\documentclass[12pt]{article}
\usepackage{graphicx}
\usepackage{amsmath}

\newcommand{\panda}{${\sf \overline{P}ANDA}$\hspace*{1ex}}

\def\pbnr{}
\def\title{New Studies of XYZ States at \panda}

\newcommand\pubnumber{\pbnr}
\newcommand\pubdate{\today}
\def\Title#1{\begin{center} {\Large #1 } \end{center}}
\def\Author#1{\begin{center}{ \sc #1} \end{center}}

\newcommand{\OnBehalf}[1]{\sbox0{#1}\ifdim\wd0=0pt
        {}% if #1 is empty
	\else
	{\\on behalf of #1}% if #1 is not empty
	\fi}
\newcommand{\SupportedBy}[1]{\sbox0{#1}\ifdim\wd0=0pt
        {}% if #1 is empty
	\else
	{\footnote{#1}}% if #1 is not empty
	\fi}
\def\Address#1{\begin{center}{ \it #1} \end{center}}

\newcommand\pubblock{\includegraphics[width=5cm]{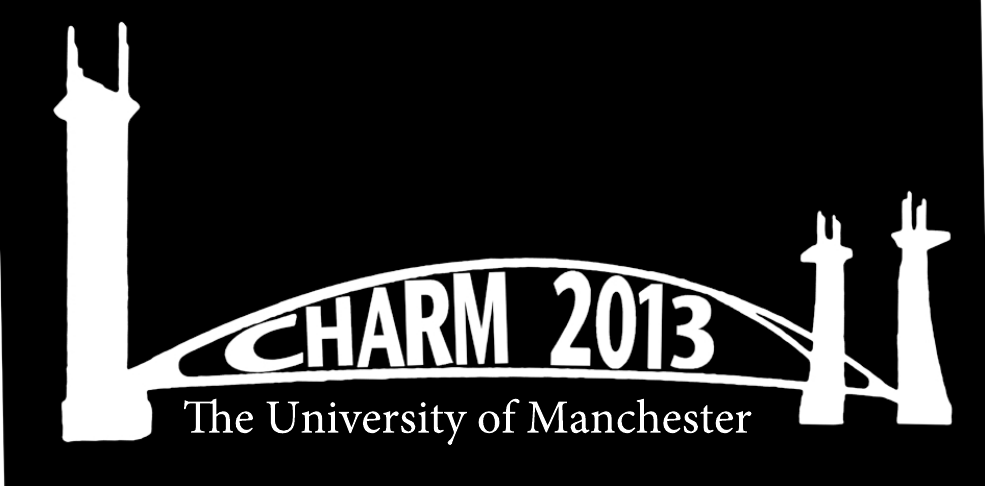}\hfill{\begin{tabular}{l} \pubnumber\\
         \pubdate  \end{tabular}}}
\newenvironment{Abstract}{\begin{quotation}  }{\end{quotation}}
\newenvironment{Presented}{\begin{quotation} \begin{center} 
             PRESENTED AT\end{center}\bigskip 
      \begin{center}\begin{large}}{\end{large}\end{center} \end{quotation}}

\def\venue{The 6$^{th}$ International Workshop on Charm Physics\\
(CHARM 2013)\\
Manchester, UK,  31 August -- 4 September, 2013}
\def\beq{\begin{equation}}
\def\eeq#1{\label{#1}\end{equation}}
\def\eeqn{\end{equation}}

\def\beqa{\begin{eqnarray}}
\def\eeqa#1{\label{#1}\end{eqnarray}}
\def\eeqan{\end{eqnarray}}

\let\bar=\overbar

\def\Dslash{\not{\hbox{\kern-4pt $D$}}}
\def\dslash{\not{\hbox{\kern-2pt $\del$}}}

\def\msb{{\bar{\ssstyle M \kern -1pt S}}}

\begin{document}

\parindent0em

\begin{titlepage}
\pubblock
\vfill
\Title{\title}
\vfill

\Author{S\"oren Lange, Martin Galuska, Simon Reiter}
\Address{II.\ Physikalisches Institut, Justus-Liebig-Universit\"at Giessen, Germany}

\Author{Elisabetta Prencipe}
\Address{Institut f\"ur Kernphysik, Forschungszentrum J\"ulich, Germany}

\Author{Stefano Spataro}
\Address{Dipartimento di Fisica, Universit\'a di Torino and INFN, Italy}

\begin{center}
on behalf of the \panda Collaboration
\end{center}

\vfill

\begin{Abstract}
\noindent
Monte Carlo simulations for charmonium(-like) states at the planned 
\panda experiment are presented,
in particular the search for the $h_c'$, the search for the $^3F_4$ state
and the search for the rare decay Y(4260)$\rightarrow$$e^+$$e^-$.
\end{Abstract}

\vfill
\begin{Presented}
\venue
\end{Presented}
\vfill
\end{titlepage}
\def\thefootnote{\fnsymbol{footnote}}
\setcounter{footnote}{0}

%%%%%%%%%%%%%%%%%%%%%%
\section{Introduction}
%%%%%%%%%%%%%%%%%%%%%%

The \panda experiment at the future 
\underline{F}acility for \underline{A}ntiproton and \underline{I}on \underline{R}esearch (FAIR),
at the GSI Helmholtz-Center, Darmstadt, Germany, is planned to start operation in 2018.
It will use a stored anti-proton beam in the 
\underline{H}igh \underline{E}nergy \underline{S}torage \underline{R}ing (HESR) 
with a momentum $p$$\leq$15~GeV/c, 
corresponding to a center-of-mass energy of $\sqrt{s}$$\leq$5.5~GeV
in a fixed target setup with e.g.\ a hydrogen pellet target. 
For charmonium(-like) states $X_{c\overline{c}}$ 
formation $p$$\overline{p}$$\rightarrow$$X_{c\overline{c}}$
or production $p$$\overline{p}$$\rightarrow$$X_{c\overline{c}}$$M$
with one or more additional mesons $M$ can be used.  
The advantage of $p$$\overline{p}$ collisions is that any quantum 
number can be formed, while in $e^+$$e^-$ collisions with one virtual photon 
only formation of $J^{PC}$=1$^{--}$ is possible.
There will be two HESR operation modes.
In the {\it high intensity mode}, using stochastic cooling,
there will be 10$^{11}$ stored anti-protons and
a beam momentum resolution of $\Delta$$p$/$p$$\simeq$10$^{-4}$.
In the {\it high resolution mode}, using electron cooling,
there will be 10$^{10}$ stored anti-protons and
a beam momentum resolution of $\Delta$$p$/$p$$\simeq$10$^{-5}$.
Additional details can be found elsewhere \cite{panda_ppr}.
In this paper, new, priorly not shown results for Monte Carlo (MC) simulations 
of charmonium(-like) states will be presented.

%%%%%%%%%%%%%%%%%%%%%%%%%%%%%%%%%%
\section{Cross sections at \panda}
%%%%%%%%%%%%%%%%%%%%%%%%%%%%%%%%%%

Cross sections in $p$$\overline{p}$ formation (as an example 
$\sigma$($p$$\overline{p}$$\rightarrow$X(3872))
can be estimated 
from measured branching fractions 
(i.e.\ ${\cal B}$(X(3872)$\rightarrow$$p$$\overline{p}$)
using the principle of detailed balance,
which is shown in Eq.~\ref{edetailed_balance}.

\begin{eqnarray}
\sigma[ p\overline{p} \rightarrow X(3872) ] & = &
\sigma_{BW}[ p\overline{p} \rightarrow X(3872) \rightarrow {\rm all}](m_{X(3872)}) \nonumber\\
& =  & \frac{(2J+1) \cdot 4\pi}{m_{X(3872)}^2 - 4 m_p^2} \cdot
\frac{{\cal B}(X(3872)\rightarrow p\overline{p}) \cdot
\overbrace{{\cal B}(X(3872)\rightarrow {\rm all})}^{=1} \cdot \Gamma_{X(3872)}^2}
{\underbrace{4(m_{X(3872)} - m_{X(3872)})^2}_{=0} + \Gamma_{X(3872)}^2} \nonumber \\
&\stackrel{(J=1)}{=} & \frac{3\cdot 4\pi}{m_{X(3872)}^2-4 m_p^2}
\cdot {\cal B}(X(3872)\rightarrow p\overline{p}) \ .
\label{edetailed_balance}
\end{eqnarray}

\begin{table*}[htb]
\begin{center}
\begin{tabular}{|l|l|l|l|l|l|}
\hline
$R$ & $J$ & $m$ $[$MeV$]$ &  $\Gamma$ $[$keV$]$ & ${\cal B}$($R$$\rightarrow$$p$$\overline{p}$) & $\sigma$($\overline{p}$$p$$\rightarrow$$R$) \\
\hline
$J$/$\psi$& 1 & 3096.916$\pm$0.011 & 92.9$\pm$2.8 & (2.17$\pm$0.07)$\times$10$^{-3}$ & 5.25$\pm$0.17~$\mu$b\\
$\psi'$ & 1 & 3686.109$^{+012}_{-014}$ & 304$\pm$9 & (2.76$\pm$0.12)$\times$10$^{-4}$ & 402$\pm$18~nb\\
$\eta_c$ & 0 & 2981.0$\pm$1.1 & (29.7$\pm$1.0)$\times$10$^3$ & (1.41$\pm$0.17)$\times$10$^{-3}$ & 1.29$\pm$0.16~$\mu$b\\
$\eta_c'$ & 0 & 3638.9$\pm$1.3 & (10$\pm$4)$\times$10$^3$ & (1.85$\pm$1.26)$\times$10$^{-4}$ & 93$\pm$63~nb\\
$\chi_{c0}$ & 0 & 3414.75$\pm$0.31 & (10.4$\pm$0.6)$\times$10$^3$ & (2.23$\pm$0.13)$\times$10$^{-4}$ & 134.1$\pm$7.8~nb\\
$h_c$ & 1 & 3525.41$\pm$0.16 & $\leq$1$\times$10$^3$ & (8.95$\pm$5.21)$\times$10$^{-4}$ & 1.47$\pm$0.86~$\mu$b\\
X(3872) & 1 & 3871.68$\pm$0.17 & $\leq$1.2$\times$10$^3$ & $\leq$5.31$\times$10$^{-4}$ & $\leq$68.0~nb\\
\hline
\end{tabular}
\end{center}
\caption{Total spin $J$, mass $m$, width $\Gamma$, branching fraction for the decay into $p$$\overline{p}$
and cross sections for production at \panda as derived by the principle of detailed balance for selected
resonances $R$.
\label{tdetailed_balance}}
\end{table*}

Tab.~\ref{tdetailed_balance} summarizes cross sections for production at \panda as derived
by the principle of detailed balance for selected resonances $R$.
For the $J$/$\psi$, the $\psi'$, the $\eta_c'$ and the $\chi_{c0}$ the branching fraction
${\cal B}$($R$$\rightarrow$$p$$\overline{p}$) was taken from \cite{pdg}.
For the $\eta_c'$,
${\cal B}$($B^+$$\rightarrow$$K^+$$R$$\rightarrow$$K^+$$p$$\overline{p}$) was taken from \cite{lhcb_ppbark}
and ${\cal B}$($B^+$$\rightarrow$$K^+$$R$) was taken from \cite{pdg}.
For the $h_c$ and the X(3872)
${\cal B}$($B^+$$\rightarrow$$K^+$$R$$\rightarrow$$K^+$$p$$\overline{p}$) was taken from \cite{lhcb_ppbark}
and the upper limit for ${\cal B}$($B^+$$\rightarrow$$K^+$$R$) was taken from \cite{pdg}.
Typical cross sections for charmonium formation at \panda are thus in the order
of 10-100~nb. 

%%%%%%%%%%%%%%%%%%%%%%%%%%%%%%%%%%%%%%%%%%%%
\section{The Quark Anti-Quark Potential}
%%%%%%%%%%%%%%%%%%%%%%%%%%%%%%%%%%%%%%%%%%%%

The static heavy quark anti-quark ($Q$$\overline{Q}$) potential of the Cornell-type
\cite{potential} can be expressed as 

\begin{equation}
V ( r ) = - \frac{4}{3} \frac{\alpha_S}{r} + k \cdot r
\label{ecornell}
\end{equation}

with a chromo-electric, Coulomb-type term and a linear confinement term. 
It predicts many of the experimentally observed charmonium and bottomonium
states up to a precision of $\simeq$1~MeV. 
Recently several new states have been observed, which fit well 
into the prediction of the Cornell-type potential, 
i.e.\ the $h_c$ \cite{hc_cleo} \cite{hc_bes3}, 
the $h_b$ and $h_b'$ \cite{hb_belle}, 
or the $\eta_b$ and $\eta_b'$ \cite{etab_belle}. 
By the mass measurements of these new states, 
a comparison of the level spacings between charmonium (mass regime 3-4 GeV) 
and bottomonium (mass regime 9-10 GeV) became available for the first time. 
As a surprising result, some of the level spacings are identical
to $\leq$1~MeV, which means a relative difference of $\geq$$10^{-4}$
compared to the mass scales \cite{soeren_paper06}. 
This important experimental observation
points to flavor independence of the potential. 
However, as already found in the 1970's \cite{quigg}, 
flavor independence is not fulfilled for a Cornell-type potential. 
Potentials, for which identical level spacings for 
charmonium and bottomonium are fulfilled, are logarithmic
potentials of the type

\begin{equation}
V ( r ) = c_1 \ln c_2 r 
\label{ecornell}
\end{equation}

with parameters $c_1$ and $c_2$.
One of the important tasks of future experiments such as \panda is the search
for additional, yet unobserved states (e.g.\ the $h_c'$ or a $^3F_4$ state), 
which could be used to obtain additional 
level spacings and further test the flavor indepedence, and possibly a logarithmic
shape of the potential. 

%%%%%%%%%%%%%%%%%%%%%%%%%%%%%%%%%%%%%%%%%%%%%%
\section{Prospects for $h_c'$ at ${\sf \overline{P}ANDA}$}
%%%%%%%%%%%%%%%%%%%%%%%%%%%%%%%%%%%%%%%%%%%%%%

\label{cpanda_hc}

The $h_c'$($n$=2, $^1P_1$) state with $J^{PC}$=0$^{-+}$ is one of the yet unobserved states,
which may be used for a test of flavor independence of the potential. 
From the Cornell-type model, it is
predicted at $m$=3934-3956~GeV \cite{potential}.
\panda is well suited for a search for the $h_c'$, 
for the following reasons:

\begin{itemize}

\item $h_c$($n$=1) was never observed in $B$ decays, 
as 0$^{-+}$$\rightarrow$0$^{-+}$1$^{+-}$ is forbidden in the factorisation limit.
In the decay $B^+$$\rightarrow$$K^+$$h_c,h_c'$ the combination of quantum numbers
would require an additional gluon connecting the $K^+$ and $h_c$ lines. 

\item The $h_c$($n$=1) ground state was observed at CLEO \cite{hc_cleo} and BESIII \cite{hc_bes3}
in the isospin violating decay $\psi'$$\rightarrow$$h_c$$\pi^0$.
However, for the $h_c'$ one would have to use the higher $\psi$(4040) 
or $\psi$(4160) resonance. As the decay would again be isospin violating, 
the branching fraction is expected to be small. In addition, the phase space 
is small, as the available kinetic energy is only $\simeq$100 or 
$\simeq$220~MeV, respectively. 

\end{itemize}

\begin{figure}[htb]
%\centerline{\includegraphics[width=0.43\textwidth,height=5.4cm]{hc1_NEW.jpg}\includegraphics[width=0.48\textwidth]{hc2_NEW.jpg}}
\centerline{\includegraphics[width=0.43\textwidth,height=4.45cm]{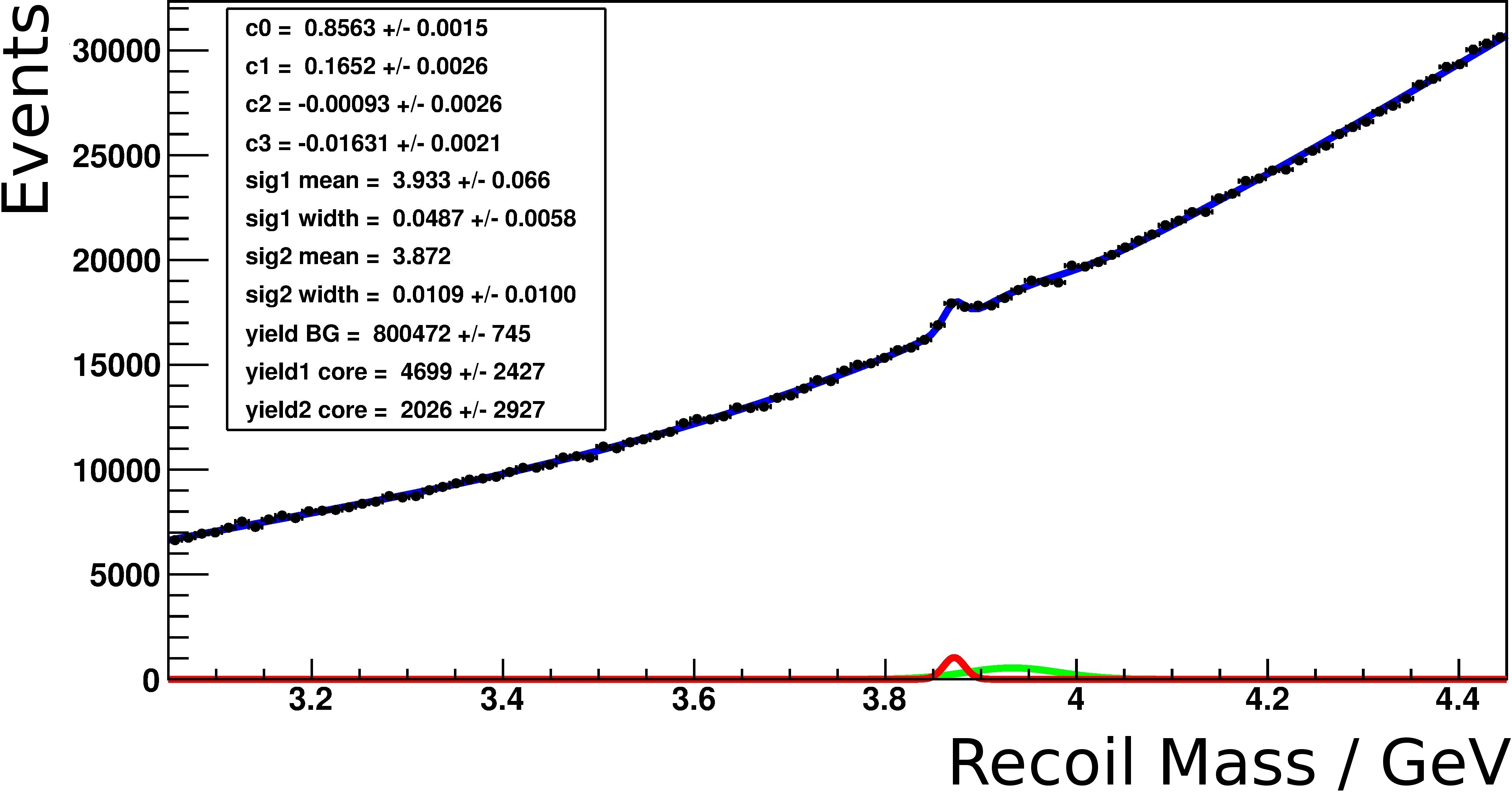}\includegraphics[width=0.48\textwidth]{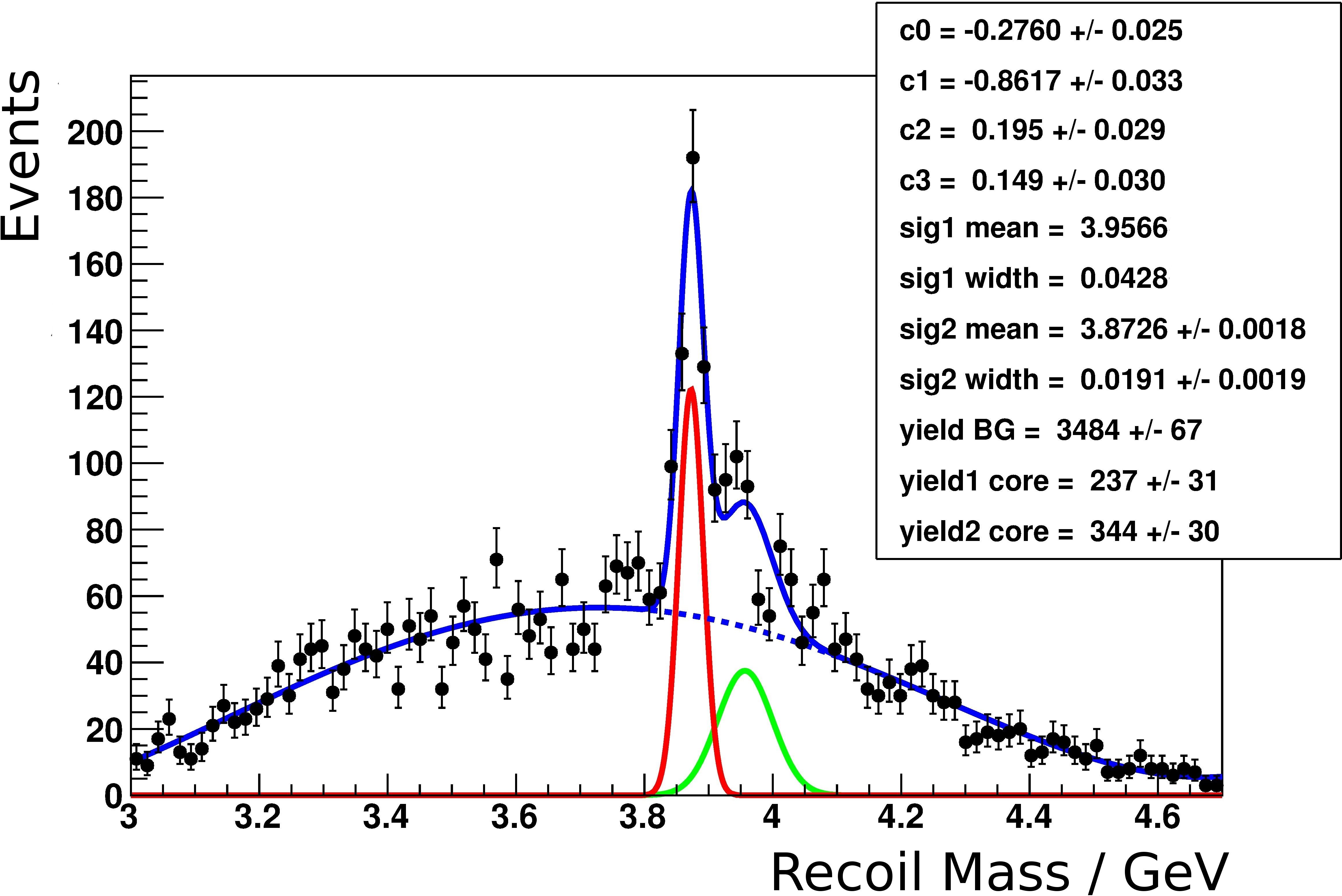}}
\caption{$\pi^+$$\pi^-$ recoil mass for a MC simulation of $p$$\overline{p}$$\rightarrow$$h_c'$$\pi^+$$\pi^-$ at \panda
for $p_{beam}$=15~GeV/c before cuts {\it (left)} and after cuts {\it (right)}. 
For details see text.\label{fhcx}}
\end{figure}

For the search for the $h_c'$ at \panda, a recoil mass technique 
provides a useful approach. MC simulations were performed
for the decay $p$$\overline{p}$$\rightarrow$($\pi^+$$\pi^-$)$_{recoil}$$h_c'$.
The advantage of this inclusive method is, that 
no knowledge on the specific decay of the $h_c'$ is required. For the simulation 
a decay $h_c'$$\rightarrow$$\eta_c$$\pi⁺$$\pi⁻$ was used on the generator
level, however with all possible $\eta_c$ decays as known from \cite{pdg}.
Fig.~\ref{fhcx} (left) shows the $\pi^+$$\pi^-$ recoil mass from a MC simulation 
$p$$\overline{p}$$\rightarrow$$h_c'$$\pi^+$$\pi^-$ at \panda \cite{bachelor_simon}.
The decay is $h_c'$$\rightarrow$$D^{0}$$\overline{D}^{0*}$
with $D^{0}$$\rightarrow$$K^-$$\pi^+$ and $\overline{D}^{0*}$$\rightarrow$anything.
The highest available anti-proton momentum of 15~GeV/c was chosen for two reasons:
{\it (a)} a higher beam momentum leads to higher reconstructable momenta and efficiencies 
of the $\pi^+$ and $\pi^-$ and {\it (b)} the inelastic $\overline{p}$$p$ cross section,
being the main source of the background $\pi^+$$\pi^-$ pairs, decreases as 
a function of beam momentum. 
An input width of $\Gamma$=87~MeV was used for the $h_c'$, consistent with 
predictions for the static potential \cite{potential}.
The assumed cross section for the signal is 4.5~nb,
corresponding to 3.9$\times$10$^4$ $h_c'$ per day produced at \panda
in the HESR high luminosity mode. 
The inelastic hadronic background cross section is $\simeq$43~mb \cite{panda_x3872_background}.
Fig.~\ref{fhcx} (left) shows the signal for 3 hours data taking and the background for 1 second of data taking,
corresponding to 2$\times$10$^7$ events, generated with the DPM model \cite{dpm}.
The number of simulated background events 
is limited by available CPU performance and will be increased in the future. 
The signal consists not only of the $h_c'$, but also of the X(3872), 
which decays into the same final state $D^{0}$$\overline{D}^{0*}$ 
and can be regarded as a reference signal for the $h_c'$.

In order to suppress the large hadronic background, three cuts were applied:
a momentum cut $p_{lab}$($\pi^{\pm}$)$>$1.2~GeV,
a vertex cut in beam direction $\Delta$$z$$\leq$0.1~cm, 
and a 3$\sigma$ cut on the invariant mass $m$($K^{\pm}$$\pi^{\mp}$)
around the nominal mass of the $D^0$. The latter cut is largely 
efficient to reduce the background. After applying the cut, 
the signal efficiency is 8.3\%, while the background efficiency
in only 1.6$\times$10$^{-5}$.
Fig.~\ref{fhcx} (right) shows the $\pi^+$$\pi^-$ recoil mass after applying the cuts. 

The above mentioned signal cross section of 4.5~nb is one main result of this analysis, 
as it represents the cross section required to achieve $S$/$\sqrt{(S+B)}$$\geq$10 in 6 weeks
of data taking with a duty factor of 50\% (for details of the calculation see \cite{bachelor_simon}).
For the plots, we assumed the relative ratio of $h_c'$ and
X(3872) to be 50\%:50\%. However, a cross section of 4.5~nb for the $h_c'$ and, as mentioned
before, an estimated cross section for the X(3872) of 50~nb would lead to a relative ratio of
9\%:91\%.

%%%%%%%%%%%%%%%%%%%%%%%%%%%%%%%%%%%%%%%%%%%%%%%%%%%%%%%%%%%%%%%%%%%%%
\section{Prospects for the $^3F_4$ state at ${\sf \overline{P}ANDA}$}
%%%%%%%%%%%%%%%%%%%%%%%%%%%%%%%%%%%%%%%%%%%%%%%%%%%%%%%%%%%%%%%%%%%%%

\label{cpanda_3F4}

One of the disadvantages of using the $h_c'$ as a test on the flavor 
independence of the potential is, that the width is as large as $\Gamma$=87~MeV.
On the other hand, the yet unobserved $^3F_4$ charmonium state is more 
appropriate due to its very narrow predicted width of 8.3~MeV \cite{potential}.
The narrow width is a consequence of the quantum numbers $J^{PC}$=4$^{++}$, 
because the decay is blocked by angular barrier. 
A transition from $L$=3 to the ground state with $L$=0 is 
suppressed by a factor (2$L$+1) with $L$=3.
Due to the same reason, the production of the $^3F_4$ state is suppressed 
in $B$ meson decays at Belle II or in radiative decays of high lying $\psi$ states at BESIII.
At \panda production of states with higher $L$ quantum numbers 
in the $\overline{p}$$p$ initial system is not suppressed,
and therefore \panda is uniquely suited for the search. 
Qualitatively $L$$\geq$10 can be achieved, but the quantitave estimates for the 
population of given $L$ values is unknown, as there are no existing measurements.
The here chosen approach for the reconstruction is the detection of a radiative cascade,
which in 3 steps changes $\Delta$$L$=1 down to the $J$/$\psi$, which then can be detected
by its decay into $e^+$$e^-$ or $\mu^+$$\mu^-$.
Tab.~\ref{tcascade} shows the parameters of the states in the radiative cascade. 

\begin{table}[hhh]
\begin{center}
\begin{tabular}{|l|l|l|l|}
\hline
1 $^3F_4$ & 1 $^3D_3$ & $\chi_{c2}$ & $J$/$\psi$\\
$J^{PC}$=4$^{++}$ & $J^{PC}$=3$^{--}$ & $J^{PC}$=2$^{++}$ & $J^{PC}$=1$^{--}$\\ 
4095~MeV & 3849~MeV & 3556~MeV & 3097~MeV \\ 
$\Gamma$=8.3~MeV & $\Gamma$=0.5~MeV & $\Gamma$=2.0~MeV & $\Gamma$=0.3~MeV \\
$E_{\gamma}$=246~MeV & $E_{\gamma}$=338~MeV & $E_{\gamma}$=413~MeV & - \\
\hline
\end{tabular}
\end{center}
\caption{Parameters of the states in the radiative cascade to search for the $^3F_4$ state at \panda.\label{tcascade}}
\end{table}

\begin{figure}[htb]
\centerline{\includegraphics[width=0.48\textwidth,height=5.3cm]{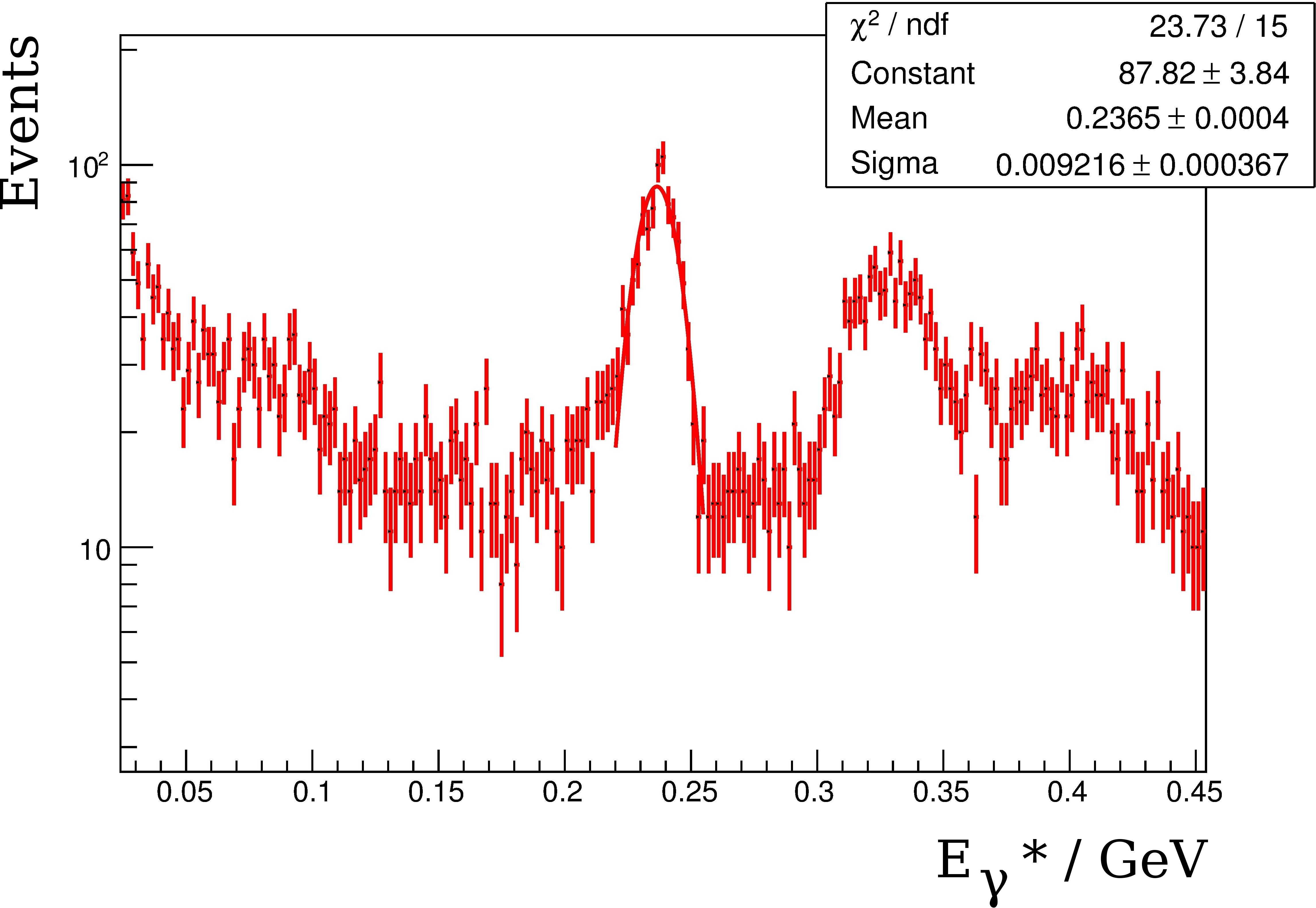}
\includegraphics[width=0.48\textwidth]{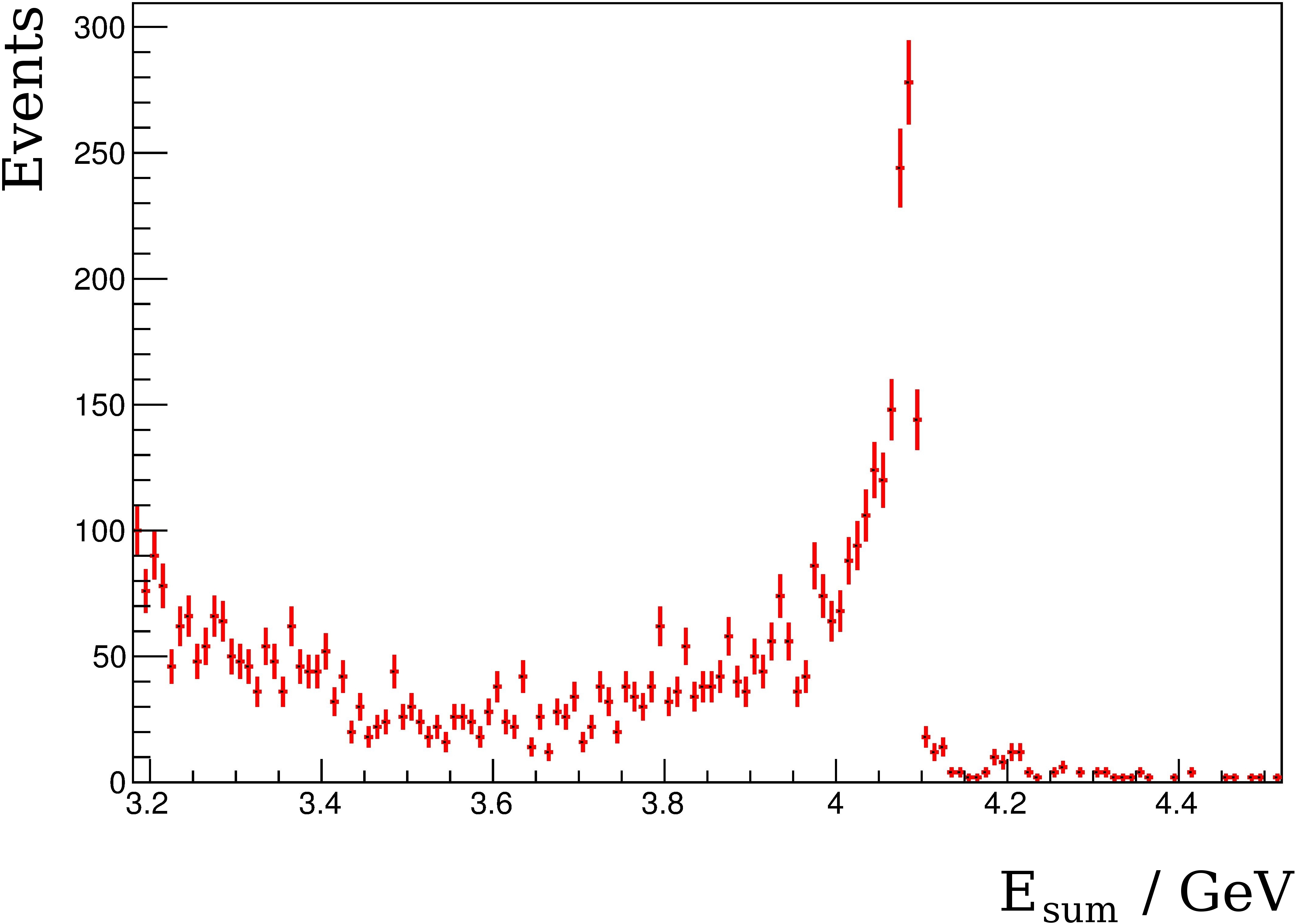}}
\caption{
MC simulation of a search for the $^3F_4$ charmonium state at \panda.
{\it Left:} $E_{\gamma}^*$ for signal events for 14 days of data taking.
{\it Right:} signal of the $^3F_4$ in the sum of the three photon energies 
$E_{\gamma 1}^*$+$E_{\gamma 2}^*$+$E_{\gamma 3}^*$ added to the nominal 
mass of the $J$/$\psi$. A four-constraint kinematical fit was applied for the three photons.\label{f3F4}}
\end{figure}

MC simulations for a search for the $^3F_4$ state at \panda were performed. 
The assumed cross section is $\sigma$($\overline{p}$$p$$\rightarrow$$^3F_4$)=10~nb.
The size of the cross section is a function of the mass of the $c$$\overline{c}$
state to be produced, and an assumption of a factor $\simeq$5 smaller cross section
than $\sigma$($\overline{p}$$p$$\rightarrow$X(3872)) is reasonable.  
The branching fraction is assumed ${\cal B}$=10\% for each of the three transitions,
corresponding to the measured value 
${\cal B}$=9.84$\pm$0.31\% \cite{pdg} for the transition $\psi'$$\rightarrow$$\gamma$$\chi_{c0}$.
Each transition was modeled with a decay from a vector meson to a vector meson and a photon
as an approximation, as $J$=2,3,4 decays are not available yet in the MC. 
The additional assumption was made that there is no polarisation. 
The search will be conducted in the HESR high luminosity mode with 8.64~$pb^{-1}$ per day.
Fig.~\ref{f3F4} (left) shows the photon energy in the center-of-mass (cms) frame $E_{\gamma}^*$ for signal events
for 14 days of data taking assuming 50\% duty factor.
The first transistion from the $^3F_4$ at 246~MeV is clearly visible and shows a photon energy 
resolution, post-boost in the cms frame, of $\sigma$($E_{\gamma}^*$)=9.2~MeV. 
Although the boost is an approximation, even the second transition with 338~MeV and the third transition at 413~MeV
are visible as well. Final state radiation in the $J$/$\psi$ decay was taken into account in the MC simulation
by using PHOTOS \cite{PHOTOS} and generates the peaking photon background at $E_{\gamma}^*$$\simeq$0~MeV. 
For suppression of this background, a cut of $E_{\gamma}^*$$\geq$150~MeV was applied in the further analysis. 
Fig.~\ref{f3F4} (right) shows the sum of the three photon energies $E_{\gamma 1}^*$+$E_{\gamma 2}^*$+$E_{\gamma 3}^*$,
where the cut was applied for each candidate photon, for 14 days of data taking assuming 50\% duty factor.
The three photons were input to a kinematical 
fit with four constraints on the total $E$, $p_x$, $p_y$ and $p_z$, 
and a cut on fit quality with $\chi^2_{fit}$$\leq$0.1 was applied. 
The nominal mass of the $J$/$\psi$ from PDG was added to adjust the mass scale. 
A narrow $^3F_4$ signal is clearly visible with a reconstructed width of $\sigma$($m$($^3F_4$))=1.2~MeV.
The background at lower masses corresponds to 43.2\% multiple candidates
due to final state radiation (as mentioned above) and Bremsstrahlung in the detector material.
The main hadronic background in this analysis is given by events with photons from light hadron ($\pi^0$, $\eta$, etc.)
decays. However, the requirement of a reconstructed $J$/$\psi$ and 3 photons with an energy cut $E_{\gamma}^*$$\geq$150~MeV
is very clean. A background suppression factor of 1.2$\times$10$^6$ was achieved for events generated with DPM,
so that the hadronic background is expected to be negligible. 

%%%%%%%%%%%%%%%%%%%%%%%%%%%%%%%%%%%%%%%%%%%%%%
\section{Prospects for the Y(4260) at ${\sf \overline{P}ANDA}$}
%%%%%%%%%%%%%%%%%%%%%%%%%%%%%%%%%%%%%%%%%%%%%%

\label{cpanda_y4260}

Another important topic for \panda is the investigation of the possible nature of the Y(4260) \cite{y4260}, 
which is often being discussed as a candidate for a [$c$$\overline{c}$$g$] hybrid state. 
From the detailed balance we can derive an estimate for an upper limit of the cross section 
$\sigma$($\overline{p}$$p$$\rightarrow$Y(4260))$\leq$4370~nb, which is unreasonably high 
due to the measured high upper limit
${\cal B}$(Y(4260)$\rightarrow$$\overline{p}$$p$)/${\cal B}$(Y(4260)$\rightarrow$$J$/$\psi$$\pi^+$$\pi^-$)$<$0.13 (90\% C.L.) 
\cite{babar_ee2ppbar}.
A better approach for obtaining a reasonable cross section is scaling from the measured 
${\cal B}$($J$/$\psi$$\rightarrow$$\overline{p}$$p$)
with the ratio of the known total widths of the $J$/$\psi$ and the Y(4260):

\begin{equation}
{\cal B} ( Y(4260) \rightarrow \overline{p} p ) 
= {\cal B} ( J/\psi \rightarrow \overline{p} p ) \times
\frac{\varGamma (J/\psi) }{\varGamma (Y(4260))} \ ,
\end{equation}

which leads to $\sigma$($\overline{p}$$p$$\rightarrow$Y(4260))=1.9$\pm$0.2 nb.
Although this is a factor $\geq$26 smaller than the cross section for the X(3872) at {${\sf \overline{P}ANDA}$},
the number of generated Y(4260) is still high. 
For the HESR high resolution mode, this correponds to 16.400 events per day,
and thus \panda may be considered a Y(4260) mini-''factory''. 
\panda is planned to achieve a peak luminosity of ${\cal L}$=2$\times$10$^{32}$~cm$^{-2}$~s$^{-1}$ 
which is only a factor $\simeq$2.7 less than the achieved peak luminosity of ${\cal L}$=5.3$\times$10$^{32}$~cm$^{-2}$~s$^{-1}$ 
on the Y(4260) resonance at BESIII. 
However, the cross section at BESIII is a factor $\simeq$30 smaller with
$\sigma$($e^+$$e^-$$\rightarrow$Y(4260))=62.9$\pm$1.9$\pm$3.7~pb \cite{zc3900_bes3}.
At Belle II, the Y(4260) will be produced in initial state radiation (ISR). 
In $B$ meson decays the Y(4260) has not been observed
so far. Based on the scaled number of observed events at Belle,
for Belle II $\simeq$30.000 ISR events are expected in an envisaged data set of 50~$ab^{-1}$,
assuming ${\cal B}$(Y(4260)$\rightarrow$$J$/$\psi$$\pi^+$$\pi^-$)=100\%.
While at Belle this would correspond to $\geq$8 years of data taking, 
at \panda only $\geq$2 days in HESR high luminosity mode would be required. 
With the very high statistics, \panda will be suited to search for rare decays 
such as Y(4260)$\rightarrow$$e^+$$e^-$. 
This decay has not been 
observed yet, although the quantum numbers of the Y(4260) with $J^{PC}$=1$^{--}$
would allow it. 
A limit on the coupling to $e^+$$e^-$ can be derived from the coupling to initial state
in $e^+$$e^-$$\rightarrow$Y(4260). However, this way only a product of a coupling to the
initial state and the coupling to the final state can be measured, as the Y(4260) must be
observed in the final state in a decay such as Y(4260)$\rightarrow$$J$/$\psi$$\pi^+$$\pi^-$.
The measured product partial width is 
${\cal B}$(Y(4260)$\rightarrow$$J$/$\psi$$\pi^+$$\pi^-$)$\times$$\Gamma$(Y(4260)$\rightarrow$$e^+$$e^−$)= 
(7.5$\pm$0.9$\pm$0.8)~eV \cite{y4260}.
Thus the partial width is of the order of eV, while the total width 
of the Y(4260) is in the order $\simeq$100 MeV, indicating the 
strong suppression of the coupling to $e^+$$^-$ by a factor $\geq$10$^7$.
Depending on the interpretation of the Y(4260), there could be several 
reasons for the suppression. In a simplified view, a suppression
could be induced by a reduced wave function at $r$=0, 
corresponding to a reduced annihilation term.
If the Y(4260) is a $[$$c\overline{c}$$g$$]$ hybrid, 
then the wave function at the origin might be reduced, 
as the minimum of the $\Pi_u$ field\footnote{The $\Pi_u$ field 
is the lowest lying gluon excitation potential, for which the 
gluon spin projected onto the quark anti-quark axis $J_G$ = 1.} 
is not at $r$=0, but at $r$$>$0
\cite{braaten_charm13} (and references therein).
If the Y(4260) is a $[$$D$$D_1$(2420)$]$ molecule, then the long-range 
part of the wave function might be enhanced and therefore, according to unitarity of the wave function, 
the short-range part at $r$$\simeq$0 is suppressed.
Tab.~\ref{tpsi2ee} shows the known branching fractions of decays 
of conventional $\psi$ charmonium states to $e^+$$e^-$ \cite{pdg}.
In order to claim a suppression, 
the question would be in fact, if the branching fraction ${\cal B}$(Y(4260)$\rightarrow$$e^+$$e^-$) is smaller than
for the $\psi$ states. 
For the reasons explained above, the measurement of such a rare decay would only be possible at 
{${\sf \overline{P}ANDA}$.
The advantage is that this would be an absolute measurement, 
not depending on the coupling to the initial state in the product branching fraction.

\begin{table}[htb]
\begin{center}
\begin{tabular}{|l|l|}
\hline
Decay & Branching fraction \\
\hline
$\psi$(3770)$\rightarrow$$e^+$$e^-$ & (9.6$\pm$0.7)$\times$10$^{-6}$ \\
$\psi$(4040)$\rightarrow$$e^+$$e^-$ & (1.07$\pm$0.16)$\times$10$^{-5}$ \\
$\psi$(4160)$\rightarrow$$e^+$$e^-$ & (8.1$\pm$0.9)$\times$10$^{-6}$ \\
$\psi$(4415)$\rightarrow$$e^+$$e^-$ & (9.4$\pm$3.2)$\times$10$^{-6}$ \\
\hline
\end{tabular}
\end{center}
\caption{Known branching fractions of decays to $e^+$$e^-$ of conventional $\psi$ charmonium states \cite{pdg}.\label{tpsi2ee}}
\end{table}

A MC simulation for $\overline{p}$$p$$\rightarrow$Y(4260)$\rightarrow$$e^+$$e^-$ was performed. 
The reconstruction efficiency turns out to be high with $\varepsilon$$>$93\% and 
only limited by acceptance and Brems\-strah\-lung, i.e.\ the $e^+$ or $e^-$ radiates one or more photons 
before being detected in the EMC, is recontructed with a wrong photon energy and therefore
may not pass the energy cuts. 
There are two main backgrounds. 
On the one hand, elastic $\overline{p}$$p$$\rightarrow$$\overline{p}$$p$
has a 2-prong signature with total energy $E$=$m$(Y(4260)). 
The cross section is high with $\sigma$=4.5$\times$10$^4$~$\mu$b, however 
a suppression technique based upon {\it (a)} the strongly peaking behavior in the polar angular distribution 
and {\it (b)} partial identification of the $\overline{p}$ annihilation in EMC
lead to a suppression of $\leq$1.3$\times$10$^{-5}$.
On the other hand, $\overline{p}$$p$$\rightarrow$$\pi^+$$\pi^-$ with a 2-prong final state 
has a high cross section as well with 4.6$\times$10$^4$~$\mu$b.
With $\pi^{\pm}$ and $e^{\pm}$ identification a suppression of $\geq$10$^6$ was achieved \cite{pid_suppression}.
Note that there can be interference between signal and background, but was not taken into account
in this analysis.

Fig.~\ref{fy4260_panda} shows the $e^+$$e^-$ invariant mass distribution 
at \panda with a beam momentum $p$=8.62323~GeV/c. 
For the Y(4260)$\rightarrow$$e^+$$e^-$ signal 3 months of data taking (50\% duty factor) 
are assumed. For the background 2$\times$10$^7$ events (generated with the DPM model),  
corresponding to one second of data taking, are shown. The number of simulated background events 
is limited by available CPU performance and will be increased in the future. 
The width of $\Gamma$=114.5$\pm$6.5~MeV was determined by a fit with a single Gaussian, 
and is consistent with the generated input width of $\Gamma$=108~MeV. 
The $J$/$\psi$ signal which is visible in Fig.~\ref{fy4260_panda}
originates from Y(4260)$\rightarrow$$J$/$\psi$$\pi^+$$\pi^-$ 
with a branching fraction of $\simeq$100\% assumed and 
subsequent $J$/$\psi$$\rightarrow$$e^+$$e^-$ with a branching fraction of 6\%.
This $J$/$\psi$ signal can be used as a reference signal
for fixing the mass scale. 
For the decay Y(4260)$\rightarrow$$e^+$$e^-$ the 
same branching fraction as $\psi$(4160)$\rightarrow$$e^+$$e^-$ in Tab.~\ref{tpsi2ee}
was assumed. 
A small contribution for Y(4260)$\rightarrow$$\psi'$$\pi^+$$\pi^-$,
which was not observed so far, assuming as a simple estimate 
${\cal B}$(Y(4260)$\rightarrow$$\psi'$$\pi^+$$\pi^-$)=${\cal B}$($\psi$(4160)$\rightarrow$$e^+$$e^-$),
was also included and is visible as the small signal for $\psi'$$\rightarrow$$e^+$$e^-$.
The fitted mass of the Y(4260) is 4.151$\pm$0.008, which is $\geq$100 MeV lower than the nomimal mass.
The reason is, that a single Gaussian, which was used here in the fit as an approximation, 
is not a proper description of the p.d.f. 
The beam momentum is adjusted to the on-resonance peak position.
Thus, the right hand side of the mass peak is only due to momentum resolution.
The left-hand side is a convolution of three effects: 
{\it (a)} a $P$-wave Breit-Wigner shape, 
{\it (b)} the momentum resolution, and  
{\it (c)} a tail from Bremsstrahlung.
This asymmetry between the left and the right hand side leads to 
the lower fitted mass. 

\begin{figure}[htb]
\centerline{\includegraphics[width=0.48\textwidth]{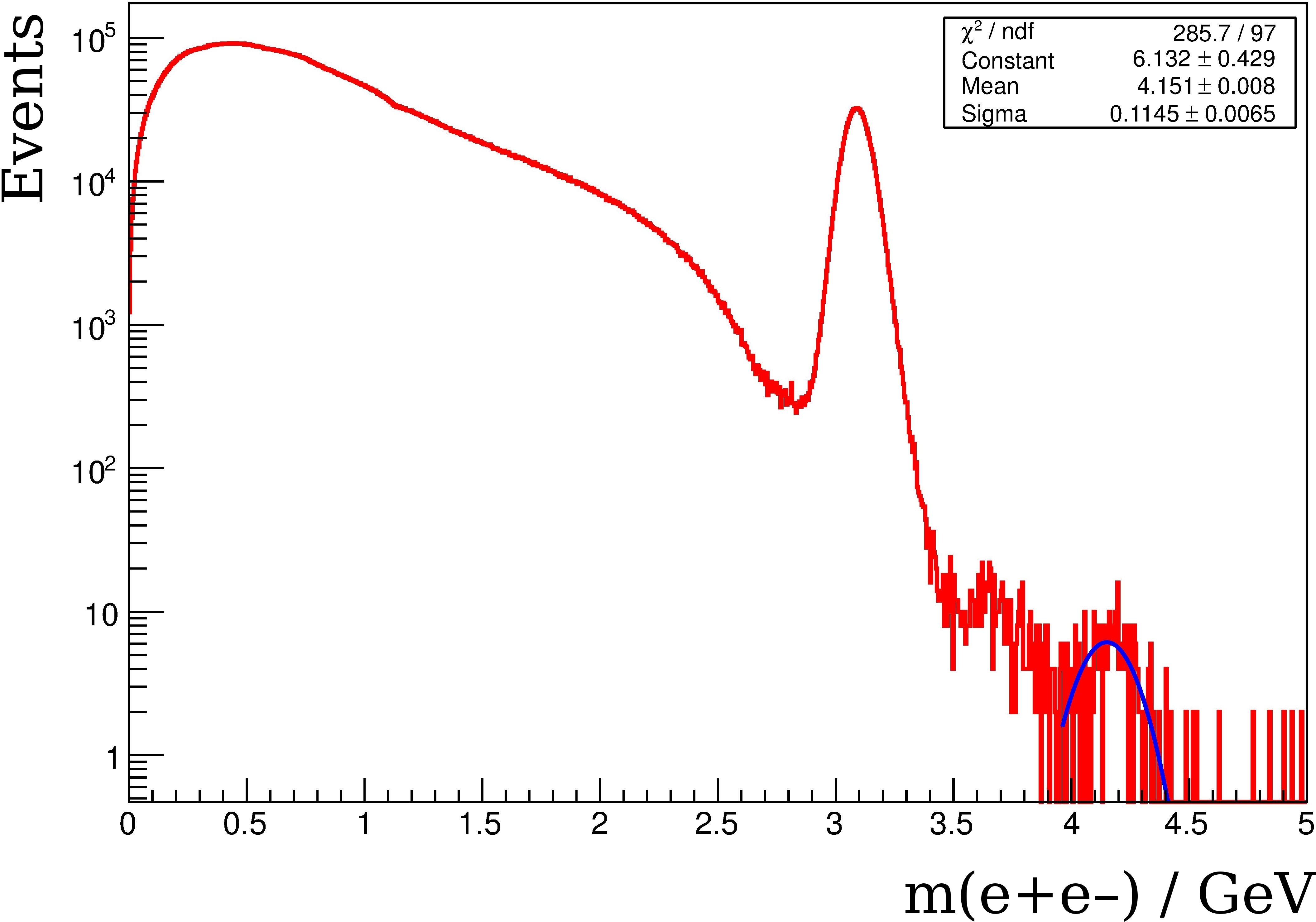}}
\caption{$e^+$$e^-$ invariant mass distribution at \panda with a beam momentum $p$=8.62323~GeV/c. 
For details see text.\label{fy4260_panda}}
\end{figure}

%%%%%%%%%%%%%%%%%
\section{Summary}
%%%%%%%%%%%%%%%%%

\panda with $p$$\overline{p}$ collisions 
is well suited for the search for high lying charmonium\-\mbox{(-like)} states, which are 
suppressed due to their quantum numbers in $B$ meson decays or radiative decays
of $\psi$ resonances. 
Expected event rates are high due to the planned high luminosity, 
e.g.\ 16.400 events with a Y(4260) per day, and thus enabling searches for rare decays
of XYZ states.

%\bibliographystyle{unsrt} %% NIEMALS INSIDE WIDETEXT !!! > thesis geht nicht mehr
%\bibliography{lange}

\end{document}